# Theoretical study of a new method for the quantum computer implementation based on self-organizing structures: the essence of new proposals and argumentation

T.M. Makhviladze[*] and M.E. Sarychev

*Institute of Physics and Technology of Russian Academy of Sciences (FTIAN), Nakhimovskii Prospect, 36/1, Moscow, 117218, Russia*

# Теоретическое исследование нового метода реализации квантовых компьютеров на основе самоорганизующихся структур: суть новых предложений и их обоснование

Т.М. Махвиладзе[*] и М.Е. Сарычев

*Физико-технологический институт Российской академии наук (ФТИАН), Нахимовский проспект, 36/1, Москва, 117218, Россия*

General key problems in relation to the application of self-organizing structures theory to the analysis of current cooperative phenomena are discussed. Particularly, we present the new results of the analysis of burning problem to form a qubit network. Recently [Proc. of the Inst. of Physics and Technol. of Russian Acad. of Sci. (Trudy FTIAN), vol. 25 (accepted for publication on Jan. 15, 2015)] we have proposed a new method for the quantum computer implementation based on self-organizing structures. Here we consider in detail the essence of new proposals and theoretical argumentation. We have shown that some self-organizing structures, both dissipative ones and thermodynamically equilibrium structures with off-diagonal long-range order, are promising to form a qubit network. This is confirmed by our results of the investigations relative to the following items: an open resonator with periodic boundaries, conditions to create specific distribution of the modes and the effect of the "field crystallization", the use of this effect to create an proficient and coherentized system of the neutral atoms traps and to achieve their precise localization; the equilibrium systems which can go, at temperatures below critical ones, into the special state with the off-diagonal long-range order (i.e. to the state with macroscopic quantum coherence), phase transition (photons condensation) in the two-level atoms interacting with radiation mode in the resonator, superradiant phase transition, the order-disorder type ferroelectrics in the cavity electromagnetic field, the advantage of the systems capable to self-organization, including the suppression of the temperature influence as the decoherence factor acting on the work quantum states of the qubits (atoms, ferroelectric molecules), and also the coherent character of the direct dipole-dipole interaction between qubits (in the case of ferroelectrics), and other items.

**Введение**

В недавних работах [1-6] нами был предложен новый способ формирования системы кубитов на основе самоорганизующихся структур [7-27].

Следует отметить, что в последние годы проводятся интенсивные фундаментальные исследования по созданию квантового компьютера в виде реального физического прибора с числом кубитов не менее $10^3$, способного решать практически значимые задачи (см. [28] и статьи по квантовым компьютерам в томах Трудов ФТИАН [29-33]). Однако, к настоящему времени построены только его весьма ограниченные варианты, причём даже в лучших реализациях обычно имеется не более двадцати связанных кубитов. То же самое фактически имеет место и в машине D-Wave, функционирующей, как заявлено в публикациях, в пределах 512 кубитов, но работающей по групповой восьми-кубитовой схеме с весьма слабыми связями между группами. Можно сказать, что образовался заметный разрыв между успехами в разработке квантовых алгоритмов и различных виртуальных приложений с одной стороны, и непосредственной физической реализации квантовых компьютеров – с другой. Ясно, что разработка теории квантовых компьютеров, квантовых алгоритмов и различных приложений, будучи весьма увлекательным разделом квантовой физики и квантовой информатики, сама по себе не может привести к созданию квантовых компьютеров – нужны новые идеи и подходы, способные дать дополнительные возможности в экспериментальном развитии этого направления.

___

[*] e-mail: tarielmakh@mail.ru

Одной из основных причин сложившейся ситуации являются трудности достижения в рамках известных технологий (твердотельные квантовые точки, сверхпроводящие элементы, атомы в оптических ловушках или ионы в магнито-электрических вакуумных ловушках, NV-центры в алмазоподобных пленках и др.) необходимой степени когерентизации достаточно большого количества кубитов. В связи с этим нам представляется, что, помимо совершенствования этих, уже получивших развитие технологий, необходим поиск таких способов организации системы кубитов, которые, являясь, возможно, более сложными и изощренными, были бы более адекватными требованиям, предъявляемым к квантово-компьютерным объектам, поскольку использовали бы изначально когерентные физические системы. Исследования такого рода систем с исходной когерентностью на наш взгляд могут оказаться перспективными для возможного преодоления ряда проблем, которые возникли при использовании уже апробированных технологий.

Задача настоящей работы состоит в том, чтобы, с целью проведения последующих исследований, привлечь внимание к иным, чем ранее применяемым, способам формирования элементной базы квантовых компьютеров, а именно - к использованию самоорганизующихся структур (см. предложение, сформулированное в работах [1-5]). Поэтому мы сочли целесообразным представить здесь результаты достаточно подробного теоретического исследования тех самоорганизующихся структур, которые могут оказаться полезными для решения проблем масштабирования, устойчивости к декогерентизации и ошибкам, проблемы редуцирования негативного влияния температуры (в особенности на ансамблевые квантовые компьютеры) и других. Основное внимание уделено изложению новых результатов, касающихся обоснования тех свойств, особенностей и характеристик самоорганизующихся систем, которые могут представлять наибольший интерес для физической реализации квантовых компьютеров.

Для пояснения предлагаемого подхода рассмотрим пример кубитов – нейтральных атомов [28], которые обычно предлагается помещать в ловушки – узлы оптических решеток, образованных стоячими волнами от лазерных источников [28]. В этом случае в качестве ловушек для размещения и локализации таких кубитов, на наш взгляд, представляет интерес использование определенных мест (точек) в полях собственных мод оптического резонатора с периодическими границами, причём расстояние между зеркалами резонатора должно подчиняться специальному условию (см. раздел 2). Возникающее при таких расстояниях специфическое пространственное распределение (самоорганизующаяся структура) собственных колебаний электромагнитного поля характеризуется появлением точек, в которых его интенсивность становится экстремально большой. Эти точки образуют в объеме резонатора двух- или трехмерную решетку и могут быть использованы для более эффективной локализации кубитов. Отметим, что генерация таких мод может осуществляться либо путем накачки специально размещенного внутри резонатора соответствующего источника поля (в оптическом диапазоне, например, активной лазерной среды), либо запуском в резонатор излучения от внешнего источника. Поле накачки может быть одночастотным и одномодовым либо многочастотным и многомодовым, вызывающем один из нелинейно-оптических процессов (например, параметрические процессы, ВКР, процессы генерации гармоник и др.). Разумеется, в указанных точках – концентраторах поля - можно располагать как один кубит, так и группы кубитов, но в последнем случае реализация квантового компьютера встретится с уже известными трудностями.

В разделе 1 изложены некоторые соображения относительно потенциальных возможностей использования диссипативных и равновесных самоорганизующихся структур для организации системы кубитов квантового компьютера..

Разделы 2, 3 посвящены теоретическому исследованию одной из таких структур - диссипативной самоорганизующейся системы собственных электромагнитных мод открытого резонатора с периодическим коэффициентом пропускания/отражения одного из зеркал. Следует указать, что для случая оптического резонатора такие моды и особенности пространственного распределения поля этих мод были предсказаны и детально исследованы теоретически А.М. Прохоровым и авторами настоящей статьи [34, 35]. Для такого распределения поля авторами был введен термин «кристаллизация поля». Впервые их существование в оптическом диапазоне было подтверждено и изучено экспериментально А.М. Прохоровым и сотрудниками при решении важной проблемы вывода лазерного излучения из резонатора с активной лазерной средой. В разделе 2 впервые представлены результаты теории открытых резонаторов с периодическими границами для полей с произвольной длиной волны, а в разделе 3 дан их анализ как диссипативных самоорганизующихся структур, образование которых происходит при определенных соотношениях между длиной резонатора, характеристиками пропускания его границ и длиной волны; основное внимание уделено тем результатам, которые полезны с точки зрения рассматриваемых здесь приложений. Подчеркнём, что возникновение особых периодических мод является следствием самоорганизации поля и имеет черты фазового перехода.

Теоретические работы [34, 35] дали толчок развитию в работах [18, 21, 36, 37] теории других самоорганизующихся систем не только открытого типа (раздел 3), но и термодинамически равновесных, которые также могут представить интерес при разработке полномасштабного квантового компьютера. В частности, к таким системам относятся термодинамические самоорганизующиеся резонаторные системы с диагональным дальним порядком по заселенностям уровней атомов (включая уровни зеемановского расщепления) и недиагональным дальним порядком по соответствующим двухуровневым переходам и резонансным им фотонам.

Например, как показано в [18, 21, 36, 37], подобными свойствами могут обладать сегнетоэлектрики или ферромагнетики в резонаторе или системы излучающих двухуровневых частиц в замкнутом резонаторе, самоорганизующиеся с одной из мод электромагнитного поля или поля квазичастиц (например, фононов) и претерпевающие фазовый переход, сопровождаемый «конденсацией фотонов» в особые коррелированные состояния - когерентные глауберовские состояния (термин «конденсация фотонов» введен авторами в [18, 21, 36, 37] и стал употребительным в теории самоорганизующихся систем рассматриваемого типа). Такого рода системы и возможные перспективы их использования для реализации квантово-компьютерных систем рассмотрены в разделе 4.

В Заключении кратко сформулированы основные результаты и выводы.

**1. О возможном использовании самоорганизующихся структур для формирования системы кубитов**

С начала 2014 года в лаборатории моделирования физико-технологических процессов микро- и наноэлектроники ФТИ РАН теоретически исследуются перспективы реализации элементной базы квантовых компьютеров на основе самоорганизующихся структур [1-5].

Предположение о том, что такие структуры могут сыграть определённую роль в организации полномасштабной системы кубитов базируется на идее о том, что можно было бы попытаться использовать присущую некоторым из них макроскопическую когерентность. Под когерентностью здесь мы имеем ввиду следующее. В самоорганизующихся структурах происходит макроскопическое упорядочение, причем тип этого упорядочения, т.е. природа так называемого «параметра порядка», для каждой конкретной системы имеет свой специфический характер. Это упорядочение связано с возникновением в определенных условиях дальнодействующих (макроскопических) корреляций между всеми элементами системы (это может быть, например, система атомов, ионов, молекул, кристаллических ячеек, дипольных моментов, спинов или система каких-либо других элементов т.д.). В ряде случаев (некоторые конкретные примеры рассмотрены ниже в разделах 2 – 4), когда система включает переменные поля той или иной природы, распространяющееся по системе, её самоорганизация может сопровождаться возникновением определенного «порядка» также и в самом поле (например, в случае электромагнитного поля – переход в глауберовское когерентное состояние). В этом случае мы и будем говорить о когерентизации системы, понимая под этим термином возникновение и развитие когерентной связи между отдельными элементами системы. Если в такой системе имеет место взаимодействие поля с какими-то ее элементами (атомными, молекулярными или др.) по переходам между какими-либо их внутренним состояниями, то в условиях самоорганизации можно говорить и о когерентности (когерентизации) самих элементов, точнее этих переходов друг относительно друга, или, что то же, о возникновении «недиагонального» дальнего порядка (подробнее см. в разделах 3,4).

Пусть теперь в качестве элементов, взаимодействующих с полем выступают объекты, претендующие сейчас на роль кубитов (различные варианты двухуровневых кубитов – ионы, атомы, молекулы, NV-центры и др.). Тогда наше предложение состоит в том, что, переводя такую систему в условия самоорганизации, можно рассчитывать, что возникшая в ней когерентность (в указанном выше смысле) будет способствовать редуцированию (блокированию) процессов декогерентизации кубитов по переходу между их рабочими квантовыми состояниями и тем самым более эффективно обеспечивать последующую многокубитовую функциональность. Отметим, что переход, по которому кубит взаимодействует с упорядочивающим полем, и переход между его рабочими (как квантового процессора) состояниями не обязаны совпадать (а, может быть, даже не должны совпадать!). Отметим также, что проблема организации когерентной связи между кубитами (обмен информацией) [28], т. е. между их рабочими переходами, представляет собой отдельную задачу, рассмотрение которой выходит за рамки настоящей статьи.

Отдельно следует сказать о случае, когда самоорганизующаяся система представляет собой само поле (здесь и далее для конкретности будем говорить об электромагнитном поле). Тогда самоорганизация может быть достигнута за счет постановки, например, периодических условий отражения на границах (см. подробнее раздел 2). Оказывается, что напряженность образующихся стоячих волн в определенных точках (местах) может стать экстремально большой. Эти точки (узлы) образуют решетку, которая могла бы представлять интерес для создания нового типа ловушек нейтральных атомов. В отличие от известных оптических решеток, которые уже используются в этом качестве, на основе предлагаемой системы возможно создание значительно более глубоких потенциальных ям. Кроме того, эта система является когерентной в том смысле, что она образуется волной от одного источника (а не от трех пар лазеров). Отметим также, что в такой системе достаточно просто решается проблема масштабирования (см. ниже). Представляется, что сочетание таких качеств может оказаться полезным при создании квантового компьютера, например, с кубитами в виде изолированных нейтральных атомов.

Как упоминалось во Введении, задел для работы по развитию предлагаемого подхода основан на предсказанных и исследованных теоретически в работах [34, 35] и затем подробно изученных экспериментально А.М. Прохоровым и сотрудниками так называемых периодических мод - собственных электромагнитных мод оптического открытого резонатора с периодическими границами, Сюда также следует отнести последующие работы [18, 21, 36, 37] по общей теории самоорганизующихся систем открытого типа, в том числе и оптических резонаторных систем, а также работы по теории равновесных самоорганизующихся систем с различными физическими механизмами возникновения дальнего порядка [6 -27].

Особую роль в свете возможного использования дополнительных механизмов для лучшей когерентизации кубитов, на наш взгляд, играют хорошо изученные специальные когерентные эффекты, наблюдаемые при взаимодействии ультракоротких лазерных импульсов с веществом (эффекты сверхизлучения [38 -43], фотонного эхо 144, 45], самоиндуцированной прозрачности [46, 47], солитонные режимы в нелинейно-оптических процессах [48,49] и др. [50, 51]). Еще раз касаясь сути подхода, изложенного в работах [1-6], следует отметить, что фактически он сводится к предложению объединить системы уже известных двухуровневых кубитов когерентной связью с помощью электромагнитного поля**.** Именно такого рода результат характерен для эффектов в полях ультракоротких (с длительностью меньшей всех времен релаксации) лазерных импульсов. В этом случае система N излучателей под действием таких импульсов переходит в состояние, из которого она дает отклик по интенсивности, пропорциональный не N, как обычно, а $N^2$, что и являлось свидетельством его когерентности. В излагаемом в данной работе подходе предлагаются системы, в которых двухуровневые кубиты в такого рода когерентном состоянии могут находиться не в течении длительности ультракороткого импульса, а сколь угодно долго. Представляется, что пребывание кубитов в таком состоянии, помимо указанной выше возможности усилить (улучшить) их когерентизацию, может способствовать и процессу инициализации, т.е. приготовлению некоторого исходного базисного состояния всех кубитов, что также необходимом для выполнения квантовых вычислений [28]. Таким образом, предлагается размещать уже физически реализованные двухуровневые однокубитовые процессоры (твердотельные квантовые точки, сверхпроводящие элементы, атомы в оптических ловушках или ионы в вакуумных ловушках, NH-центры в алмазоподобных пленках и др.) или те, которые появятся в будущем, в новых условиях, включаюших дополнительный когерентизирующий фактор (резонаторное электромагнитное поле) и приводящих к самоорганизации всей системы.

Для лучшего понимания нового подхода кратко остановимся на технологии формирования системы кубитов в виде нейтральных атомов, помещенных в оптические ловушки. Такие системы активно исследуются для создания кубитов квантовых компьютеров, и при традиционной подходе пути решения проблемы масштабирования таких устройств, по-видимому, в настоящее время можно считать трудно разрешимыми. В то же время, перспективной может оказаться возможность реализации таких ловушек с помощью самоорганизующейся системы мод оптического резонатора с периодическими границами, то есть, создание особого рода диссипативной (самоорганизующейся) структуры внутри резонаторного поля, представляющего собой оптическую решетку с экстремально большой интенсивностью в узлах. Такая решетка и может быть использована для создания более эффективных ловушек нейтральных атомов, чем разработанные ранее. Заметим, что ловушки, образованные концентраторами поля, характеризуются наличием экстремально глубоких двумерных потенциальных ям, возможно, снимая в значительной степени проблему охлаждения атомов, захваченных в эти ловушки (более сложной экспериментально является создание ловушек с трехмерной потенциальной ямой – для этого необходимо «закрыть» систему по оси, перпендикулярной оси резонатора; в этом случае мы фактически будем иметь волновод специального вида).

В разделе 2 дано подробное исследование собственных полей открытого резонатора. В определенных условиях они имеют структуру когерентной решетки (подобно решетке в открытом оптическом резонаторе) с экстремально усиленной напряженностью поля в пучностях, что, по нашему мнению, делает перспективными их использование в качестве ловушек для атомных кубитов. Для исследования этих полей в разделе 2 в рамках общей теории открытых (диссипативных) структур развито описание электромагнитного поля внутри и вне резонатора с периодическими границами. Целью развития теории являлся анализ стационарных распределений поля для различных периодических коэффициентов отражения излучения от одного из зеркал резонатора и его геометрических параметров. Необходимо было также детально исследовать собственные моды внутри резонатора; проанализировать особенности пространственного распределения амплитуд собственных мод в зависимости от параметров коэффициента отражения и расстояния между зеркалами резонатора, исследовать эффекты «кристаллизации» фотонов. Наконец, исследовано излучение системы в дальней зоне (вне резонатора), анализ которого может позволить проводить диагностику поля внутрирезонаторных ловушек с размещенными в них кубитами.

Как уже отмечалось во введении, в качестве использования базиса для построения квантовых компьютеров интерес могут представить термодинамические самоорганизующиеся системы. Поэтому далее (раздел 4) развит последовательный подход для расчета равновесной термодинамики системы двухуровневых кубитов (атомы, спины в магнитном поле, дипольные ячейки в сегнетоэлектриках и т.д.), взаимодействующих с резонансными фотонами. Используя этот подход и включая в гамильтониан дополнительно фононные моды, проведены детальные расчеты и подробно исследована термодинамика системы двухуровневых объектов, взаимодействующих между собой через мягкую фононную моду (по типу сегнетоэлектриков) и с фотонами. Аналитически исследованы основные термодинамические величины таких систем, включая параметры порядка (диагональный и недиагональный), а также температурные зависимости указанных параметров порядка. На основании этого анализа сделаны выводы о возможности фазового перехода в упорядоченное состояние, сопровождающегося когерентизацией как системы двухуровневых объектов-кубитов, так и поля; даны оценки температур такого перехода.

Следует отметить, что необходимо тщательное теоретическое исследование предлагаемого подхода к организации системы кубитов и соответствующие детальные эксперименты; в данной статье мы хотели рассмотреть, в первую очередь нетривиальные свойства самоорганизующихся структур, которые не стоит игнорировать, учитывая трудности с перспективами физической реализации идеи квантовых компьютеров.

## 2. Теория открытых резонаторов с периодическими границами

В настоящем разделе приводятся результаты теоретического исследования открытых резонаторов с периодическими границами (ОРПГ). На основе решения краевой задачи уравнений электродинамики проведено исследование собственных колебаний ОРПГ, вычислены поля внутри и вне резонатора для различных периодических структур, выяснена физическая природа возникновения периодических мод и указан способ их диагностики, что необходимо при реализации расположения кубитов в точках – концентраторах поля.

### 2.1. Основные соотношения

Рассмотрим систему, состоящую из двух бесконечных плоских зеркал, поверхности которых определяются соотношениями: $z=\pm l$, $-\infty < x, y < +\infty$. Пусть одно из зеркал (при $z=-l$) имеет не зависящий от координат амплитудный коэффициент отражения $T_1=1-\beta_1$, а второе — зависящий от $x$ и $y$ коэффициент $T_2(x, y)$. Зеркала будем считать достаточно тонкими так, чтобы можно было пренебречь изменением полей при их распространении внутри зеркал. Взяв зависимость полей от времени в виде $\exp(-i\omega t)$, где $\omega=ck$ ($k$ — волновое число), будем исходить из скалярного волнового уравнения для какой-либо компоненты $W$ электромагнитного поля. Можно показать, что в условиях слабого поглощения оно сводится к параболическому уравнению, решение которого удобно представить в виде [52]

$$W(\xi, \eta, \zeta) = \Phi_1((\xi, \eta, \zeta) \exp(2ikl\zeta) - (-1)^q \Phi_2(\xi, \eta, -\zeta) \exp(-2ikl\zeta),$$

$$\xi=(k/2l)^{1/2}x, \quad \eta=(k/2l)^{1/2}y, \quad \zeta=z/2l,$$

где $\Phi_1(\xi, \eta, \zeta)$ определяет поле волны, набегающей на зеркало $z=l$, $\Phi_2(\xi, \eta, -\zeta)$ — вызванное им дифракционное поле, $q$ — целое число. Для поперечных компонент электрического поля граничные условия имеют вид

$$\Phi_1(\xi, \eta, -½) = \exp(i\chi)T_1\Phi_2(\xi, \eta, ½),$$

$$\Phi_2(\xi, \eta, -½) = \exp(i\chi)T_2(\xi, \eta) \Phi_1(\xi, \eta, ½), \qquad (1)$$

где $\chi=2kl-\pi q$. Соотношения (1) отвечают граничным условиям на поверхности зеркал при малых углах падения волны. Для факторизуемого коэффициента отражения $T_2(\xi,\eta) =$

$T_2{}^a(\xi)\,T_2{}^b(\eta)$ трехмерная задача сводится к двумерной [35], если положить

$$\Phi_i = \Phi_i^a(\xi, \zeta)\,\Phi_i^b(\eta, \zeta), \quad i = 1,2.$$

Поэтому в дальнейшем будем рассматривать уравнения

$$\frac{\partial^2 \Phi_i(\xi,\ \zeta)}{\partial \xi^2} + 2i\frac{\partial \Phi_i(\xi,\ \zeta)}{\partial \zeta} = 0 \qquad (2)$$

с граничными условиями

$$\Phi_1(\xi, -½) = \exp(i\chi)T_1\Phi_2(\xi, ½),$$

$$\Phi_2(\xi, -½) = \exp(i\chi)T_2(\xi)\,\Phi_1(\xi, ½),$$

причем

$$W(\xi, \zeta) = \Phi_1(\xi, \zeta)\exp(2ikl\zeta) - (-1)^q\,\Phi_2(\xi, -\zeta)\exp(-2ikl\zeta).$$

Нетрудно получить интегральные уравнения

$$f_1(\xi) = \exp(2i\chi)T_1\int_{-\infty}^{\infty} K_1(\xi,\xi')f_1(\xi')d\xi' \qquad (3)$$

$$f_2(\xi) = \exp(2i\chi)\,T_1T_2(\xi)\int_{-\infty}^{\infty} K_2(\xi,\xi')f_2(\xi')d\xi' \qquad (4)$$

определяющие падающее и отраженное поля при z = -*l*:

$$f_1(\xi) = \Phi_1(\xi, -½), \quad f_2(\xi) = \Phi_2(\xi, -½).$$

Ядра уравнений (3), (4) имеют вид

$$K_1(\xi,\xi') = \int_{-\infty}^{\infty} \Gamma(\xi - \xi'', 1)T_2(\xi'')\Gamma(\xi'' - \xi', 1)d\xi'',$$

$$K_2(\xi,\xi') = \int_{-\infty}^{\infty} \Gamma(\xi - \xi'', 1)\Gamma(\xi'' - \xi', 1)d\xi'',$$

где $\Gamma(\xi-\xi', \zeta - \zeta')$ — функция Грина уравнения (2), а значения $\chi$ должны находиться как собственные числа уравнений (3), (4) и давать частотный спектр собственных колебаний резонатора.

### 2.2. ОРПГ с синусоидальным амплитудным коэффициентом отражения

Рассмотрим решение уравнений (3), (4) для синусоидальной решетки:

$$T_2(\xi) = 1 - \beta_2(1+ m \cos \alpha\xi), \quad \alpha = (2\pi/p)(2l/k)^{1/2}, \tag{5}$$

где p — период решетки, $0 \leq \beta_2 \leq \frac{1}{2}$, $0 \leq m \leq 1$. Будем искать решение (4) в виде

$$f_2(\xi) = \exp(is_j\xi) \sum_{n=-\infty}^{\infty} A_n(\alpha) \exp(in\alpha\xi),$$

Тогда

$$\tfrac{1}{2}(1-\beta_1)\beta_2 m A_{n-1}\exp[2i\alpha^2 n - i(\alpha^2 - 2\alpha s_j)] - A_n\{(1-\beta_1)(1-\beta_2) -$$

$$-\exp[i(s_j+n\alpha)^2 - i\bar{\chi}]\} + \tfrac{1}{2}(1-\beta_1)\beta_2 m A_{n+1}\exp[-2i\alpha^2 n -$$

$$-i(\alpha^2 + 2\alpha s_j)] = 0, \quad \bar{\chi} = 2\chi. \tag{6}$$

Корни уравнения $\det\{A\}=0$, где $\{A\}$ - матрица коэффициентов $A_n$ системы (6), определяют зависимость $\bar{\chi}=\bar{\chi}(\alpha, s_j)$, т. е. собственные частоты ОРПГ.

При $\beta_2=0$ получаем $A_n=0$ $(n \neq 0)$ и

$$\bar{\chi} = \bar{\chi}_0 = s_j^2 - 4\pi j + i \ln(1-\beta_1), \quad j=0, \pm 1,\ldots, \tag{7а}$$

где $S_j$ - произвольное действительное число, а $f_2(\xi)$ имеет вид бегущей или стоячей волны [52]:

$$f_2^{(0)}(\xi) = \exp(is_j\xi), \quad f_2^{(0)}(\xi) = \cos(s_j\xi), \quad f_2^{(0)}(\xi) = \sin(s_j\xi). \tag{7б}$$

Так как $\beta_2 \leq 1/2$, решение (4) можно получить, применяя теорию возмущений по $\beta_2$:

$$f_2(\xi) = f_2^{(0)}(\xi) + \beta_2 f_2^{(1)}(\xi) + \beta_2^2 f_2^{(2)}(\xi) + \ldots, \quad \bar{\chi} = \bar{\chi}_0 + \beta_2 \mu_1 + \beta_2^2 \mu_2 + \ldots \tag{8}$$

Выбирая $f_2^{(0)}(\xi) = \cos(s_j\xi)$, получим

$$f_2^{(1)}(\xi) = A_1^{(1)}\cos(\alpha+s_j)\xi + A_{-1}^{(1)}\cos(\alpha-s_j)\xi, \quad \mu_1 = -i/2,$$

где

$$A_{\pm 1}^{(1)} = -\tfrac{1}{2}m\{1-\exp[-i(\alpha^2 \pm 2\alpha s_j)]\}^{-1}.$$

Такую же структуру имеют функции $f_2^{(n)}(\xi)$, полученные в высших порядках теории возмущений [35]. Аналогичным способом строится и решение уравнения (3).

Существование решения (8) показывает, что в области справедливости теории возмущений в системе имеются собственные колебания при любых $\lambda$, $p$ и $L=2l$. Они представляют собой решения (7) для резонатора с идеально отражающим зеркалом $z=l$ плюс добавки, обусловленные периодической модуляцией $T_2(\xi)$.

Структура коэффициентов разложения $f_2^{(n)}$ по $\cos(s_j \pm n\alpha)\xi$ показывает, что в окрестностях значений $\alpha$ и $s_j$, определяемых из равенств

$$\alpha^2 + 2\alpha s_j = 2\pi n_1, \quad \alpha^2 - 2\alpha s_j = 2\pi n_2, \quad n_{1,2} = 0, \pm 1, \pm 2, \ldots, \tag{9}$$

теория возмущений перестает быть справедливой. Отметим, что в силу условия $(\operatorname{Re} k)l \gg 1$ все величины в (9) можно считать действительными. Из (9) имеем

$$\alpha^2 = \pi(n_1 + n_2), \quad s_j^2 = s_{j,n1}^2 = \pi(n_1 - n_2)^2 / 4(n_1 + n_2) \quad (n_1 > -n_2). \tag{10}$$

При заданной величине $\alpha^2$, удовлетворяющей первому из условий (10), произвольным может являться только одно из чисел $n_1$ и $n_2$. Поэтому соответствующие значения $s_j$ далее нумеруются индексом $n_1$. Приведем выражения для полей при выполнении (9). Здесь возможны два случая:

а) $(n_1 + n_2)$ — нечетное. Тогда

$$\alpha^2 = \pi(2r+1), \quad s_{j,n1}^2 = \pi[2(n_1-r)-1]^2/4(2r+1)$$

($r$ — целое число) и

$$W(\xi,\zeta) = C_1 \sum_{n=-\infty}^{\infty} \exp[-i(s_{j,n1} + 2n\alpha)^2/4] \cos((s_{j,n1} + 2n\alpha)\xi) \times$$

$$\times \left\{ \left(\frac{1-\beta_1}{1-\beta_2}\right)^{1/2} \exp[-\tfrac{1}{2}\zeta \ln(1-\beta_1)(1-\beta_2)] \times \right.$$

$$\times \exp[-i\pi(2j-q+2nn_1+n(2n-1)(2r+1)\zeta] -$$

$$- (-1)^{q+n} \exp[\tfrac{1}{2}\zeta \ln(1-\beta_1)(1-\beta_2)] \times$$

$$\left. \times \exp[i\pi(2j-q) + 2nn_1 + n(2n-1)(2r+1)\zeta] \right\}, \tag{11}$$

где $C_1$ — произвольная константа;

б) $(n_t + n_2)$ — четное. Тогда

$$\alpha^2 = 2\pi r, \quad s_{j,n1}^2 = \pi(n_1-r)^2/2r$$

и поле в ближней зоне зеркала $\zeta = 1/2$ имеет вид

$$W\left(\xi, \frac{1}{2}\right) = \frac{\beta_2(1-\beta_1)^{1/4}}{(1-\beta_2)^{3/4}} \exp\left[-\frac{i}{4}(s_{j,n1}^2 - 2\pi q')\right] \cdot$$

$$\cdot \cos(s_{j,n1}\xi) \left\{ C_1 \sum_{n=-\infty}^{\infty} (-1)^n \cos 2n\alpha\xi + 2iC_2 \sum_{n=1}^{\infty} (-1)^{n-1} \sin(2n-1)\alpha\xi \right\},$$

$$q' = q - 2j,$$

$C_1$ $C_2$ — произвольные константы. Частотный спектр в обоих случаях определяется соотношением

$$\bar{\chi} = s_{j,n1}^2 - 4\pi j + i \ln(1-\beta_1)(1-\beta_2) \qquad (12)$$

Структура полей в ближней зоне становится ясной, если использовать следующее представление: в случае а)

$$W\left(\xi, \frac{1}{2}\right) = C_1 \cdot \frac{\pi \beta_2 (1-\beta_1)^{1/4}}{\alpha (1-\beta_2)^{1/4}} \exp\left[-\frac{i}{4}(s_{j,n1}^2 - 2\pi q')\right] \cdot$$
$$\cdot \cos(s_{j,n1}\xi) \sum_{n=-\infty}^{\infty} \delta\left[\xi - (2n-1)\frac{\pi}{2\alpha}\right], \qquad (13)$$

в случае б)

$$W\left(\xi, \frac{1}{2}\right) = \frac{\pi \beta_2 (1-\beta_1)^{1/4}}{\alpha (1-\beta_2)^{1/4}} \exp\left[-\frac{i}{4}(s_{j,n1}^2 - 2\pi q')\right] \cdot$$
$$\cos(s_{j,n1}\xi) \sum_{n=-\infty}^{\infty} [C_1 + i(-1)^n C_2] \delta\left[\xi - (2n-1)\frac{\pi}{2\alpha}\right]. \qquad (14)$$

Из (13), (14) следует, что поле в ближней зоне границы $\varsigma = 1/2$ имеет максимумы в точках перегиба коэффициента отражения (5). Нетрудно показать, что поле в ближней зоне зеркала $\varsigma = -1/2$ имеет ту же структуру, но в случае а) его максимумы смещены на $p/4$ относительно максимумов поля при $\varsigma = 1/2$.

Рассмотрим подробнее условия (9). Из них следует, что

$$L = 2l = (p^2/2\lambda)(n_1 + n_2).$$

Таким образом, в ОРПГ при заданном $L$ возможны собственные колебания вида (13), (14) с длинами волн, кратными $p^2/2L$. При других длинах волн колебания такого вида невозможны, и собственные поля определяются согласно (8).

В рассмотренных выше случаях а) $n_1+n_2=2r+1$ и б) $n_1+n_2=2r$ расстояния $L$ соответственно равны

$$L_1 = L_0(2r-1)/4, \quad L_2 = L_0 r/2 \quad (L_0 = 2p^2/\lambda, \quad r=1, 2,\ldots).$$

Следует отметить, что, как впервые было показано Рэлеем (см., например, [14]), на расстояниях, равных целому числу $L_0$, происходит воспроизведение бесконечно протяженной пассивной дифракционной решетки при ее облучении плоской волной. Таким образом, возникновение особых резонаторных свойств в случаях, когда $L=L_{1,2}$, связано с воспроизведением волнового фронта на границах. Это согласуется с трактовкой резонаторных систем, данной в [53].

Согласно (9), (12)

$$k = \frac{\pi}{2l} q' + \frac{1}{4l} s_{j,n1}^2 + \frac{i}{4l} \ln(1-\beta_1)(1-\beta_2).$$

(15)

В силу (Re $k$)$l \gg 1$ при любом $\beta_2$ мнимая часть $k$ пренебрежимо мала по сравнению с действительной. Из (11) следует, что эти решения можно считать полями, имеющими z-составляющую волнового вектора:

$$k_z = \frac{\pi}{2l} q' + \frac{i}{4l} \ln(1-\beta_1)(1-\beta_2)$$

и x-составляющую $k_x = s_{j,n1} (k/2l)^{1/2}$. Поля (13), (14) можно трактовать как волны, распространяющиеся с «квазиимпульсом» $k_x$ по дискретным «узлам» $x_n = p(2n-1)/4$. В частности, при $s_{j,n1} = 0$ колебания в этих «узлах» происходят синфазно. Выражение (15) теперь можно представить в виде $k = k_z + k_x^2/2\kappa$, т. е. как разложение $k = (k_x^2 + k_z^2)^{1/2}$ по малому параметру $|\kappa_x/\kappa_z|$. Поэтому указанное представление справедливо при $s_{j,n1}^2 \ll \pi q'$, причем $q' \gg 1$, так как (Re $k$)$l \approx |k|l \gg 1$. Отсюда можно заключить, что углы падения фронта волны на зеркала $\vartheta = |\kappa_x/\kappa_z| \ll \vartheta = |k_x/k_z| \ll 1$, ется с введением не зависящего от $\vartheta$ коэффициента отражения диэлектрической поверхности. Возможные типы собственных колебаний (13), (14) должны идентифицироваться двумя целыми числами $q'$ и $n_1$. Добротность этих колебаний

$$Q_{q'} = -\pi q' / \ln(1-\beta_1)(1-\beta_2)$$

### 2.3. ОРПГ со ступенчатым коэффициентом отражения

Пусть

$$T_2(\xi) = 1 - \beta_2 + \beta_2 \Pi(\xi), \quad 0 \leq \beta_2 < 1, \quad \textbf{(16)}$$

где

$$\Pi(\xi) = \begin{cases} 0, & \text{если } |\xi - n\bar{p}| \leq \bar{\tau}/2, \quad n = 0, \pm 1, \ldots, \\ \gamma, & \text{если } \bar{\tau}/2 < \xi - n\bar{p} < \bar{p} - \bar{\tau}/2, \end{cases}$$

$\bar{p} = p(k/2l)^{1/2}$ - безразмерный период, $\bar{\tau} = \tau(k/2l)^{1/2}$, $\sigma = (p-\tau)/\tau$ - скважность решетки. Построив решения (3), (4) в виде ряда по степеням $\beta_2$ аналогично (8), можно показать, что при условиях (9) собственные колебания перестают определяться этим рядом.

Рассмотрим далее колебания, возникающие при выполнении соотношений (9). Можно показать [35], что в этом случае собственные числа $\bar{\chi}$ определяются выражениями (12). Возможные типы собственных колебаний, частоты и добротности определяются из формулы (15), в которой следует изменить интервал возможных значений $\beta_2$: $0 \leq \beta_2 < 1$.

Выражение для поля на периодической границе имеет вид

$$W\left(\xi, \frac{1}{2}\right) = A \frac{\beta_2 (1-\beta_1)^{1/4}}{(1-\beta_2)^{3/4}} \exp\left[-\frac{i}{4}(s_{j,n1}^2 - 2\pi q')\right] \cos(s_{j,n1}\xi) G(\xi) \quad (17а)$$

где

$$G(\xi) = \begin{cases} g(\xi), & \text{если } |\xi - n\bar{p}| \leqslant \bar{\tau}/2, \\ 0, & \text{если } \bar{\tau}/2 < \xi - n\bar{p} < \bar{p} - \bar{\tau}/2, \end{cases} \quad (18)$$

$g(\xi)$ — произвольная функция. Поведение периодической с периодом $\bar{p}$ амплитудной части поля на зеркале $\varsigma = -1/2$ зависит от значения $s_{j,n1}$:

$$W\left(\xi, -\frac{1}{2}\right) = -\frac{A\beta_1 \exp[-i(s_{j,n1}^2 + 2\pi q')/4]}{[(1-\beta_1)(1-\beta_2)]^{1/4}} \cdot \quad (17б)$$
$$\cdot \cos(s_{j,n1}\xi) \cdot \begin{cases} G(\xi), & n_1 - \text{четное}, \\ G(\xi + \pi/\alpha), & n_1 - \text{нечетное}. \end{cases}$$

Существенно, что решения (17) существуют только, если $(n_1 + n_2)$ — четное

$$L = L_2 = L_0 r/2, \quad s_{j,n1}^2 = \pi(n_1 - r)^2/2r. \quad (19)$$

Следует отметить, что выражение (18) дает только общую структуру функции $G(\xi)$, но не ее конкретный вид. Найденное собственное колебание ОРПГ с $L=L_2$, соответствующее фиксированным индексам $n_1$ и $q'$, может иметь различную зависимость от поперечной координаты $\xi$. Конкретный вид такой моды может определяться условиями возбуждения ОРПГ.

Из (17) видно, что поля в ближней зоне периодического зеркала отличны от нуля там, где $T_2(\xi)$ имеет минимумы — «дырки» и, наоборот, равны нулю на тех участках, где $T_2(\xi)$ максимален. Легко показать, что существуют собственные колебания с противоположной структурой поля в ближней зоне:

$$\overline{W}\left(\xi, \frac{1}{2}\right) = \overline{A} \frac{\bar{\beta}_2(1-\beta_1)^{1/4}}{(1-\bar{\beta}_2)^{3/4}} \exp\left[-\frac{i}{4}(s_{j,n1}^2 - 2\pi q') \cos(s_{j,n1}\xi) \overline{G}(\xi) \right. \quad (20а)$$

$$\overline{W}\left(\xi, -\frac{1}{2}\right) = -\frac{\overline{A}\beta_1}{[(1-\beta_1)(1-\bar{\beta}_2)]^{1/4}} \exp\left[-\frac{i}{4}(s_{j,n1}^2 + 2\pi q')\right] \quad (20б)$$
$$\cdot \cos(s_{j,n1}\xi) \cdot \begin{cases} \overline{G}(\xi), & n_1 - \text{четное}, \\ \overline{G}(\xi + \pi/\alpha), & n_1 - \text{нечетное}. \end{cases}$$

Здесь

$$\bar{\beta}_2 = \beta_2(1-\gamma), \quad (n_1 + n_2) - \text{четное},$$
$$\overline{G}(\xi) = \begin{cases} 0, & \text{если } |\xi - n\bar{p}| \leqslant \bar{\tau}/2, \\ \bar{g}(\xi), & \text{если } \bar{\tau}/2 < \xi - n\bar{p} < \bar{p} - \bar{\tau}/2, \end{cases}$$

причем собственные числа $\bar{\chi}$ даются формулой

$$\bar{\chi} = s_{j,n1}^2 - 4\pi j + i \ln(1-\beta_1)(1-\beta_2+\beta_2\gamma).$$

На зеркале $z = -l$ распределения полей с нечетным $n_1$, смещены на $p/2$ относительно распределения полей с четным $n_1$: на периодическом зеркале такое смещение отсутствует. Максимумы интенсивности на периодическом зеркале попадают на участки с большими значениями коэффициента отражения. Добротности собственных колебаний (20)

$$\bar{Q}_{q'} = -\pi q'/\ln(1-\beta_1)(1-\beta_2+\beta_2\gamma)$$

превышают добротности колебаний (17). Моды (17), (20) в дальнейшем будем называть периодическими модами.

Общий случай ОРПГ с периодическим коэффициентом пропускания $T_2(\xi)$, когда $\Pi(\xi)$ в формуле (16) – произвольная функция, может быть рассмотрен с использованием тех же методов, которые были изложены выше; схема соответствующего теоретического подхода была указана ещё в статье [35]. В виду громоздкости получаемых результатов они будут опубликованы позже в отдельной статье. Отметим здесь только, что периодические моды в общем случае периодического коэффициента пропускания обладают целым рядом интересных особенностей, поскольку возможно варьирование дополнительных параметров, описывающих геометрию коэффициента пропускания. Эти особенности, связанные с пространственной структурой периодических мод, могут дать новые дополнительные и нетривиальные возможности при использовании ОРПГ в технологии формирования системы кубитов (см. раздел 3).

Для перехода к трехмерному случаю достаточно заметить, что если $T_2(x)$ и $T_2(y)$ определяются одинаковой зависимостью вида (16), то при $L=L_2$ полная индексация мод должна задаваться тремя числами $q'$, $n_1$ и $m_1$, причем $k_y = s_{j,m1}(k/2l)^{1/2}$, где

$$s_{j,m1}^2 = \pi(m_1-m_2)^2/4(m_1+m_2), \quad m_2 = 2\lambda L/p^2 - m_1 \quad (m_1+m_2=n_1+n_2).$$

Если периоды $T_2(x)$ и $T_2(y)$ различны и равны $p_x$ и $p_y$, то система имеет периодические моды при $L = p_x^2(n_1+n_2)/2\lambda$, причем $n_1+n_2=(p_y/p_x)^2(m_1+m_2)$, где $(m_1+m_2)$ — четное.

Следует отметить, что метод рассмотрения ОРПГ как задачи на собственные колебания является приближенным. Метод тем точнее, чем больше добротность колебаний. Математическим выражением сделанного приближения является пренебрежение мнимостью $k$ в условиях (9).

### 2.4. Поле в дальней зоне. Диагностика мест размещения кубитов.

Если при $z=l$ волновой фронт ограничен апертурой $|x|\leq a$, поле в дальней зоне в приближении Фраунгофера определяется обычной формулой:

$$F(x, z) = -\frac{i}{\lambda z} \exp\left(i\frac{2\pi}{\lambda} z\right) \int_{-a}^{a} W(x_1, l) \exp\left(-i\frac{2\pi}{\lambda z} x x_1\right) dx_1 \exp\left(i\frac{\pi x^2}{\lambda z}\right),$$

(21)

где $W(x_1, l)$ — поле в ближней зоне. Выражение (21) описывает дифракционную картину при $y = 0$. Двумерная картина в случае одномерной решетки $T_2(x, y) = T_2(x)$ получается умножением (21) на $\text{sinc}(by/\lambda z)$, где $\text{sinc } u = \sin(\pi u)/\pi u$, $(-b, b)$ — апертура по оси $y$. В случае двумерной решетки полная дифракционная картина получается умножением (21) на

$$F(y, z) = \exp\left(i\frac{\pi}{\lambda z} y^2\right) \int_{-b}^{b} W(y_1, l) \exp\left(-\frac{2\pi}{\lambda z} y y_1\right) dy_1,$$

где $W(y, l)$ — поле в ближней зоне, полученное решением двумерной задачи с $T_2 = T_2(y)$. Имея в виду этот переход к трехмерному случаю, ниже ограничимся вычислением дифракционного поля при $y=0$.

В случае синусоидального коэффициента (5) главные максимумы интенсивности имеют одинаковую высоту и располагаются под углами

$$\theta_n = \left[\frac{4\pi}{p} n \pm s_{j,n1}\left(\frac{\pi}{\lambda l}\right)^{1/2}\right] \frac{\lambda}{2\pi}$$

Рассмотрим случай ОРПГ со ступенчатым коэффициентом отражения (16) при возбуждении периодической моды (20). Считая для определенности, что $\overline{G}(x)$ — четная функция:

$$\overline{G}(x) = \sum_{n=0}^{\infty} b_n \cos(2\pi n x/p)$$

для интенсивности излучения $I = |F(x, z)|^2$ при условии $p \ll a$ имеем

$$I = \frac{\overline{A}^2 \overline{\beta}_2^2 (1-\beta_1)^{1/2} a^2}{(1-\overline{\beta}_2)^{3/2} \lambda^2 z^2} \sum_{m=-\infty}^{\infty} b_n'^2 \left\{\text{sinc}^2\left[\left(s_{j,n1}\left(\frac{\pi}{\lambda l}\right)^{1/2} - \frac{2\pi x}{\lambda z} + \frac{2\pi}{p} n\right)\frac{a}{\pi}\right] + \right.$$

(22)

$$\left. + \text{sinc}^2\left[\left(-s_{j,n1}\left(\frac{\pi}{\lambda l}\right)^{1/2} - \frac{2\pi x}{\lambda z} + \frac{2\pi}{p} n\right)\frac{a}{\pi}\right]\right\}, \quad b_0' = 2b_0, \quad b_n' = b_n \quad (n \neq 0)$$

мов интенсивности, располагающихся по направлениям

$$\theta_n = \left[\frac{2\pi}{p} n \pm s_{j,n1}\left(\frac{\pi}{\lambda l}\right)^{1/2}\right] \frac{\lambda}{2\pi}.$$

Угол между соседними максимумами равен $\lambda/p$, а их ширина, определяемая по первым нулям $\text{sinc } u$, $\delta\theta = \lambda/a$. Зависимость приведенной интенсивности

$$\bar{I} = \frac{z^2 \lambda^2 (1-\bar{\beta}_2)^{3/2}}{\bar{A}^2 \bar{\beta}_2^2 a^2 (1-\beta_1)^{1/2}} I$$

от $\bar{x} = \rho x/\lambda z$ есть функция с резкими максимумами при

$$\bar{x}_n = n + p\ s_{j,n1}(\pi\lambda l)^{-1/2}/2,$$

значения которой при $\bar{x} = \bar{x}_n$ равны $b_n^{'2}$, т. е. $b_n^2$ — огибающая этих максимумов. Так как $\lim b_n = 0$ при $n \to \infty$, огибающая выделяет некоторое количество максимумов $n^*$ около нулевого. Насколько эффективно происходит это выделение, зависит от $p$, $\sigma$ и конкретного вида $\bar{g}(x)$. Нетрудно видеть, что наиболее узко направленное излучение соответствует однородному возбуждению периодической моды $\bar{g}(x) = $const, когда $b_n$ определяются только скважностью решетки:

$$b_0 = \frac{\sigma}{1+\sigma}, \qquad b_n = \frac{2}{1+\sigma}\ \text{sinc}\ \frac{n}{1+\sigma} \qquad (n \neq 0). \qquad (23)$$

Можно считать, что $b_n/b_0 < 1$ при $|n\sigma/(1+\sigma)| \geq 1$, т. е. для $|n| \geq [1 + \sigma^{-1}]$ (символ [...] означает целую часть числа). Таким образом, $n^* = 2[1+\sigma^{-1}] - 1$, если $1+\sigma^{-1} - [1+\sigma^{-1}] \ll 1$ и

$n^* = 2[1+\sigma^{--1}] + 1$, если $1+\sigma^{-1} - [1+\sigma^{-1}]$ не слишком мало. Число $n^* = 1$ при $\sigma \gg 1$, и эффективная угловая расходимость $\delta\theta$ излучения определяется дифракцией на апертуре: $\delta\theta \sim \lambda/a$; диаграмма направленности излучения ОРПГ представляет собой один узкий лепесток.

Полученные в этом разделе соотношения дают прямой и достаточно простой способ диагностики мест размещения кубитов при использовании самоорганизующихся резонаторных систем.

### 3. ОРПГ как самоорганизующаяся диссипативная структура

Сразу же после разработки теории открытого оптического резонатора с периодическими границами-[34, 35] было показано [18], что в пространственном распределении поля в таком резонаторе может возникать специфическое упорядочение, носящее характер самоорганизуюшейся диссипативной структуры. Напомним, что в настоящее время в ряде областей физики, химии и биологии широко изучены и интенсивно развиваются представления о диссипативных структурах [6-10, 12,13, 17-18] - упорядоченных временных или пространственных конфигурациях, которые возникают в открытых системах вдали от равновесного состояния. Для случая оптического ОРПГ в работах [34,35] было показано, что оптическая диссипативная структура проявляется в виде особого пространственного распределения амплитуды поля (эффект

"кристаллизации" поля). Пользуясь результатами раздела 2 настоящей статьи и имея в виду исследование возможности применения ОРПГ для целей построения квантовых компьютеров, рассмотрим здесь кратко возможность генерирования самоорганизующейся диссипативной структуры в открытом (но не обязательно оптическом) резонаторе с периодическими границами.

Рассмотрим собственные колебания оптического поля в системе, состоящей из двух бесконечных пластин, поверхности которых определяются соотношениями: z=±l, $-\infty < x, y < +\infty$. Пусть одна *из* пластин (при z = -l) является идеально отражающей, а вторая (z = l) имеет зависящий от *x*, y амплитудный коэффициент отражения T(x', y) (дифракционная решетка). Взяв зависимость поля от времени в виде exp(-iωt) (ω = ck, k- волновое число), будем исходить из скалярного волнового уравнения для какой-либо компоненты W электромагнитного поля $(\Delta + k^2)$ W = 0, решение которого представим в виде

$$W(x,y,z) = \Phi_1(x,y,z) \exp(ikz) - (-1)^q \Phi_2(x,y,-z) \exp(-ikz),$$

где $\Phi_1(x,y,z)$ определяет поле набегающей волны, $\Phi_2(x,y,-z)$ - вызванное им отраженное поле (q - целое число). Считая далее для простоты, что коэффициент отражения, а следовательно, и поле в резонаторе не зависят от y: T = T(x), $\Phi_1 = \Phi_1(x,z)$, $\Phi_2^* = \Phi_2(x, -z)$, и переходя к безразмерным координатам $\xi = x\sqrt{k/2l}$, $\zeta = z/2l$, в силу обычно выполняемого условия l Re k ≫ 1 получим:

$$\frac{\partial^2 \Phi_r}{\partial \xi^2} + 2i \frac{\partial \Phi_r}{\partial \zeta} = 0, \ r=1,2. \tag{24}$$

Компоненты поля при ζ = - ½ удовлетворяют граничному условию W(ξ,-1/2) = 0 (идеально проводящая пластина); другие граничные условия имеют вид

$$\begin{aligned} f_1(\xi) &\equiv \Phi_1(\xi,-1/2) = \exp(i\chi) \Phi_2(\xi,1/2), \\ f_2(\xi) &\equiv \Phi_2(\xi,-1/2) = \exp(i\chi) T(\xi)\Phi_1(\xi,1/2), \end{aligned} \tag{25}$$

где $\chi = 2kl - \pi q$. Используя функцию Грина уравнения (24) и граничные условия (25), легко показать аналогично [18]. что задача о собственных колебаниях сводится к решению системы линейных однородных интегральных уравнений для $f_1(\xi)$ и $f_2(\xi)$.

В случае синусоидального коэффициента отражения

$$T(\xi) = 1 - \beta(1 + m \cos \alpha\xi),$$

где $\alpha = (2\pi/p)\sqrt{2l/k}$, p - период решетки, $0 \leq \beta \leq 1/2$, $0 \leq m \leq 1$, решения граничной задачи (25) при произвольных значениях длины волны λ, периода решетки p и расстояния L = 2l являются периодическими по ξ колебаниями вида:

$$f_r(\xi) = \exp(iS_j\,\xi) \sum_{n=-\infty}^{\infty} A_n^{(r)}(\alpha, S_j)\exp(in\alpha\xi)$$

где величины $S_j$ имеют смысл x-компоненты волнового вектора поля и принимают любые действительные значения, а коэффициенты $A_n^{(r)}$ представляются в виде рядов по степеням параметра $\beta\,m \leq 1/2$. Поля $f_r(\xi)$ представляют собой моды вида $\exp(iS_j\,\xi)$ с малыми, пропорциональными величинам $(\beta m)^n$, добавками высших гармоник вида $\exp[i(S_j \pm n\alpha)\xi]$.

Существенно иная картина колебаний возникает, если длина системы L равна

$$L = (2\rho^2/\lambda)(n_1 + n_2) \equiv \bar{L}(n_1, n_2),\ n_{1,2} = 0, \pm 1, \pm 2, \ldots\ (n_1 > -n_2), \qquad (26)$$

а $s_j$ определяется выражениями

$$s_j = (\pi/4)(n_1 - n_2)^2 / (n_1 + n_2) \equiv \bar{S} \qquad (27)$$

Тогда коэффициенты $A_n^{(r)}$ перестают определяться разложениями по степеням параметра $\beta m$ и поле в резонаторе претерпевает качественную перестройку. Например, если $(n_1 + n_2)$-нечетное, то на периодической границе поле имеет вид

$$W(\xi, 1/2) = C\,[\pi\beta/\alpha(1-\beta)^{3/4}]\exp\{(-i/4)[\bar{S}^2 - 2\pi(q-2j)]\}\exp(i\bar{S}\xi) \times \qquad (28а)$$

$$\times \sum_{n=-\infty}^{\infty} \delta[\xi - (2n-1)\pi/2\alpha]$$

а в центральной плоскости $\zeta = 0$

$$W(\xi, 0) = C(\pi/2\alpha)\exp(-i\bar{S}2/4)\exp(i\bar{S}\xi)\sum_{n=-\infty}^{\infty}[(1/\sqrt{(1-\beta)} - \exp(i\pi q)) -$$

$$- i(-1)^n\exp(-i\alpha\bar{S})((1/\sqrt{(1-\beta)} + \exp(i\pi q))]\delta[\xi-\pi(2n+1)/4\alpha], \qquad (28б)$$

где C - произвольная постоянная, j - целое число.

Важная особенность колебаний (28) - периодических мод - состоит в том, что они не зависят от амплитуды m периодической модуляции коэффициента отражения $T(\xi)$. В частности, в пределе m→0, когда модуляция $T(\xi)$ "выключается", колебания (28) тем не менее сохраняются. Другая особенность полей (28) заключается в том, что в пределе β→0, когда граница $\zeta = 1/2$ становится идеально отражающей ($T(\xi)=1$) и, соответственно, поле на этой границе обращается в нуль *(W(ξ, 1/2) = 0)*, собственные колебания в виде периодических мод опять-таки сохраняются. В частности, в центральной плоскости они описываются выражением (28б) с β =0.

Таким образом, при выполнении условий (26), (27) рассматриваемая система аномально чувствительна к периодической модуляции коэффициента отражения на границе, даже если эта модуляция исчезающе мала. Другими словами, отсутствует непрерывность собственных колебаний по амплитуде m коэффициента отражения $T(\xi)$.

Это указывает на то, что при $L = \bar{L}(n_1, n_2)$, $S_j = \bar{S}$ в системе происходит перестройка поля в состояние с периодическими модами (28). Отметим, что аналогичная ситуация имеет место при рассмотрении фазового перехода в ферромагнетиках, когда для выделения определенного направления намагниченности, следуя методу квазисредних Боголюбова, в задачу вводят затравочное магнитное поле, которое затем устремляют к нулю (при температурах ниже критической ферромагнетик аномально чувствителен к малым изменениям магнитного поля около нулевого значения - магнитная восприимчивость имеет особенность). В рассматриваемой здесь системе для получения колебаний (28) необходимо ввести затравочную диссипацию энергии, промодулированную по $\xi$ и задаваемую параметрами m, β.

Подчеркнем, что указанный фазовый переход, сопровождающийся образованием новой структуры поля (28), осуществляется в открытой системе, так как он связан с модуляцией коэффициента отражения, т.е. с пропусканием излучения через границу и диссипацией энергии. Для получения стационарной структуры (28) в систему должно поступать излучение извне.

Из сказанного ясно, что для данной открытой системы поля (28) имеют смысл диссипативных структур, формирующихся при изменении длины системы L. Отметим, что аналогичную роль размер системы может играть при образовании диссипативных структур в нелинейных химических реакциях с диффузией (см., например, модель "брюсселятора" [6]). Интересно, что структура поля (28а) - сумма дельта-функций - похожа на распределение плотности в одномерном идеальном кристалле с узлами в точках $x_n = (p/4)(2n - 1)$. В этом смысле возникновение диссипативной структуры (28) можно назвать эффектом "кристаллизации" поля (термин, введенный в [18] при рассмотрении оптического ОРПГ). Сказанное особенно наглядно, если $s_{j,n1} = 0$; тогда поле, имея сложный вид в объёме резонатора, стягивается в "узлы" на периодической границе и в плоскости $\zeta = 0$. Отметим ещё, что такими же свойствами обладают собственные поля для периодических коэффициентов отражения общего вида. Однако в этом случае "узлы", в которых локализуются поля, размыты и имеют размер, равный τ (см. формулу (16)). Меняя этот параметр, то есть скважность периодической границы, и длину резонатора, можно регулировать расположение узлов и пучностей самоорганизующейся диссипативной структуры электромагнитного поля.

Отметим ещё, что рассмотренная система описывается линейными интегральными уравнениями относительно граничных значений полей $f_{1,2}(\xi)$. Следовательно, в отличие, от известных случаев образования диссипативных структур [6, 9, 19, 25], описывающихся нелинейными дифференциальными уравнениями, здесь, как и для оптических ОРПГ [18] показана возможность возникновения диссипативной структуры в однокомпонентной линейной системе. При выполнении критических условий (26), (27) эта структура (28) представляет собой сумму определенным образом сфазированных гармоник, т.е. при $L = \bar{L}(n_1, n_2)$ имеет место кооперативное поведение гармоник с $S_j = \bar{S}$. Возникновение диссипативной структуры такого рода связано, с тем, что граничные поля $f_1$ и $f_2$ в каждой точке $\xi$ определяются распределением этих полей по всей системе, т.е. с нелокальным характером обратной связи..

## 4. Равновесные самоорганизующиеся структуры с недиагональным дальним порядком

Прежде, чем переходить к рассмотрению свойств самоорганизующихся равновесных структур, приведём несколько общих соображений о влиянии электромагнитных полей на термодинамические свойства вещества. К настоящему времени эта область исследований охватывает широкий круг веществ, которые могут находиться в разных агрегатных состояниях, и многие типы излучений. Несомненно, наиболее интересным здесь является изучение фазовых переходов.

Проведенные к настоящему времени экспериментальные и теоретические работы по фазовым превращениям, либо идущим под действием световых нолей, либо не требующим внешних воздействий, но связанным с электромагнитными переходами между дискретными энергетическими уровнями среды, можно разделить на три основных направления.

Первое — самое обширное — направление связано с изучением фазовых переходов, которые вызваны тепловым воздействием потоков излучения, вводимых в вещество. Значительный прогресс в этой области был обусловлен использованием интенсивных световых источников — оптических квантовых генераторов, способных давать направленные потоки излучения большой мощности.

Второе направление изучает фазовые переходы, которые не связаны с тепловым эффектом, а вызваны перестройкой взаимодействия между частицами среды вследствие изменения тех или иных характеристик частиц (населенностей уровней, дипольных моментов, сил притяжения и отталкивания между частицами и т. д.) под действием внешних полей. Наглядным примером здесь служит фазовый переход газ — жидкость, идущий под действием света [54, 55]. Вследствие воздействия резонансного излучения, переводящего частицы (атомы или молекулы) с основного энергетического уровня на какой-либо возбужденный, может происходить заметное изменение параметров уравнения состояния системы газ — жидкость, а значит, и критических параметров вещества. Оно обусловлено тем, что ван-дер-ваальсовский потенциал притяжения между возбужденными светом атомами в ряде случаев может сильно отличаться от потенциала взаимодействия частиц, находящихся в основных состояниях. Перестройка сил межмолекулярного взаимодействия под действием света приводит к зависимости вида кривой сосуществования газообразной и жидкой фаз от интенсивности света. В результате фазовый переход газ — жидкость может индуцироваться или подавляться излучением (в зависимости от того, является ли сила притяжения между возбужденными частицами больше или меньше силы притяжения между невозбужденными частицами). Изменение разностей населенностей энергетических уровней (зон) или перестройка межчастичного взаимодействия под действием излучения могут приводить к влиянию излучения на целый ряд фазовых переходов: нормальный металл — сверхпроводник, металл — диэлектрик, на фазовый переход в сегнетоэлектрическое состояние, фазовые переходы в жидких кристаллах и др. Интересны также примеры проявления коллективных эффектов в условиях воздействия на среду стационарного когерентного светового импульса и в химически активных средах: явление бистабильности оптического пропускания [56-58] и ряд эффектов в когерентной фотохимии [23, 59, 60] (ср. с результатами работ [7, 19, 20, 24-27], где исследованы эффекты самоорганизации в химических автокаталитических и конвективных системах в отсутствие поля: формирование странных аттракторов, бистабильности и бифуркации разного типа и др.).

Оба эти направления относятся к неравновесным фазовым переходам, сопровождающихся установлением стационарного состояния («потокового» равновесия) в системе среда — внешнее поле.

Третье направление, которому посвящен настоящий раздел, изучает фазовые переходы, когда среда и излучение находятся в тепловом равновесии. Его развитие началось с появления работы Хеппа и Либа [11]. В этой работе было показано, что система энергетических спинов, взаимодействующих с линейным (полевым) осциллятором, при определенном условии на величину константы взаимодействия (его называют условием сильной связи) испытывает фазовый переход второго рода по температуре. Рассмотрение, проводившееся в [11], касалось системы $N$ двухуровневых частиц, взаимодействующих с одной модой электромагнитного поля в резонаторе в условиях теплового равновесия, т.е. резонатор служит также тепловым резервуаром системы. Такая система описывается гамильтонианом (так называемая модель Дике):

$$\mathcal{H} = \varepsilon a^+ a + \frac{1}{2}\varepsilon \sum_{j=1}^{N} \sigma_j^z + \frac{\lambda}{N^{1/2}} \sum_{j=1}^{N} (a^+ \sigma_j^- + a\sigma_j^+), \qquad (29)$$

где ε — расстояние между энергетическими уровнями частицы; λ — амплитуда перехода под действием электромагнитного поля (в [11] λ считалась действительной); $\sigma_j^z$, $\sigma_j^\pm$ — матрицы Паули, описывающие *j*-ю частицу; $a, a^+$ — бозе-операторы рождения и уничтожения фотонов резонансной моды. Гамильтониан (29) описывает в двухуровневом приближении систему атомов или молекул, взаимодействующих с электромагнитным полем (представление энергетического спина). Он является точным при описании взаимодействующей с электромагнитным полем системы реальных спинов, находящихся в постоянном магнитном поле. В условиях сильной связи

$$\lambda^2 > \varepsilon^2 \tag{30}$$

рассматриваемая система испытывает фазовый переход второго рода. Когда температура становится ниже критической

$$T_c = \varepsilon \; (2k \; \text{arcth} \; (\varepsilon^2/\lambda^2))^{-1},$$

система переходит в новое равновесное состояние, характеризующееся сильной фазовой корреляцией между отдельными частицами. При этом в резонаторе возникает макроскопически заполненная мода электромагнитного поля.

Под этим понимается следующее: $\lim_{N \to \infty} \frac{\langle a^+ a \rangle}{N} \neq 0$ ($\langle ... \rangle$ - термодинамическое среднее), т. е. число фотонов в системе пропорционально числу частиц. Отметим, что при планковской функции распределения фотонов, которая имеет место при $T > T_c$ (или во всем температурном интервале, если условие сильной связи (30) не выполняется), указанный предел обращается в нуль. Далее, следуя работе [21], в случае осуществления фазового перехода с макроскопическим заполнением фотонной моды ($<a^+a> \sim N$), будем употреблять термин «конденсация фотонов». Отметим, что данный эффект не относится к типу так называемых квантовых фазовых переходов (КФП) (см. например [61]). КФП прелставляют собой переход между макроскопическими квантовыми состояниями системы при околонулевой абсолютной температуре, происходящий за счет квантовых флуктуаций в системе. В то же время рассматриваемая нами «конденсация фотонов» достигается в результате фазового перехода 2-го рода по температуре, т.е. за счет тепловых флуктуаций.

Ввиду ограниченности объёма статьи, мы приведем далее только обзор тех наших результатов, посвященных равновесному фазовому переходу в системе среда — поле, которые имеют непосредственное отношение к проблеме организации двухуровневых кубитов. Подробное изложение теории и её результатов, первые из которых были получены ещё в работах [15, 16, 63, 64, 65, 21, 62] сразу после появления статьи [11], будет представлено в последующих наших публикациях. Здесь же мы кратко перечислим в основном только самые последние результаты, необходимые для понимания физического смысла фазового перехода в системе излучатели — поле, рассмотрению природы конденсата и смысла условия сильной связи (30). Отметим здесь ещё, что о наблюдении фазового перехода в системе Дике (называемого также сверхизлучательным фазовым переходом или фазовым переходом Хеппа-Либа) впервые сообщалось, насколько нам известно, в работе [66].

Основные черты возникающего в условиях сильной связи коррелированного состояния излучателей, как оказалось, можно выяснить благодаря использованию когерентных состояний углового момента и исследованию энергетического спектра системы. Следует отметить, что именно фазовой корреляцией молекул, вызванной взаимодействием через общее поле излучения, обусловлен целый ряд когерентных эффектов, называемых также коллективными или кооперативными, которые широко изучаются в нелинейной оптике и квантовой электронике. Примерами таких эффектов служат сверхизлучение [15,38-43, 50], сверхизлучательное комбинационное рассеяние [67, 39, 51] световое эхо [45], световое эхо в нелинейных процессах [44, 68-72], эффект самоиндуцированной прозрачности [46, 47], солитоны ВКР [48, 49, 73], нутационный эффект [74-78], рамановские биения [75] и др. [58]. Эти эффекты реализуются в неравновесных условиях, создаваемых, как правило, когерентной импульсной накачкой, причем условием их проявления — условием когерентности взаимодействия излучения с веществом — служит малость характерного времени радиационного взаимодействия (длительности накачки) по сравнению с временами продольной и поперечной релаксации в среде. При достаточно больших временах наблюдения релаксационные процессы приводят к состоянию термодинамического равновесия, разрушая радиационную корреляцию излучателей. Тем более интересен случай фазового перехода в замкнутой системе излучатели — поле, когда, эффекты фазовой корреляции реализуются в условиях термодинамического равновесия [11, 15, 16, 63, 62, 64, 65].

Нами было показано, что при $T < T_c$ основное состояние такой системы представляет собой глауберовское когерентное состояние по полю и специфическое коррелированное состояние — когерентное состояние момента — по суммарному энергетическому спину частиц. Таким образом, когерентизация системы, которая желательна для формирования сети кубитов, достигается в условиях сильной связи. Наши исследования показывают, что это условие (30) является весьма жестким. Поэтому потребовалось специальное исследование для того, чтобы изучить возможности его выполнения или попытаться его обойти.

Мы, в частности, провели исследование термодинамических свойств системы многоуровневых частиц, способных взаимодействовать с несколькими резонансными модами поля. Такое обобщение теории Хеппа и Либа оказалось целесообразным, поскольку многоуровневые системы обнаруживают целый ряд важных особенностей, представляющих интерес для формирования системы кубитов (см. раздел 1 и публикации [1-6]). Одна из них — возможность когерентизации кубитов по их рабочему (задействованному в квантовых вычислениях) переходу за счет достижения состояния конденсации фотонов по более низкоэнергетическим переходам, располагающихся между двумя рабочими уровнями кубитов. Существенно, что, во-первых, для получения эффекта когерентизации не требуется выполнения условия сильной связи (30) по самому рабочему высокоэнергетическому переходу и, во-вторых, не требуется, чтобы резонатор был настроен на частоту этого перехода. Таким образом, этот переход остается «свободным» (неравновесным) для управления состоянием кубита в вычислительном процессе. Наконец, немало важно, что, как следует из соотношения (30), по низкоэнергетическим переходам условие сильной связи достигается легче, чем по переходам с более высокой энергией. Такие возможности в случае многоуровневых частиц связаны с тем, что рассматриваемый фазовый переход относится к классу фазовых переходов с недиагональным дальним порядком. Если при $T < T_c$ оказывается отличными от нуля недиагональные элементы матрицы плотности но соответствующим низкоэнергетическим переходам, то автоматически возникает недиагональный дальний порядок по «замыкающему» высокоэнергетическому переходу, причем для этого не нужны ни условие сильной связи, ни сама полевая мода, резонансная этому переходу (разумеется, эта мода должна «конденсироваться», если её замкнуть резонатором, играющим одновременно роль теплового резервуара).

Другая возможность, способная привести к конденсации фотонов без выполнения условия сильной связи, впервые обсуждалась в работах [21], где предлагалось использовать фазовые превращения вещества, сопровождающиеся возникновением недиагонального дальнего порядка. Например, возникающий при сегнетоэлектрическом (ферромагнитном) переходе недиагональный дальний порядок, как будет показано ниже, позволяет получать макроскопическое заполнение резонаторной моды электромагнитного поля без выполнения условия сильной связи по полю. Далее излагаются последние результаты, относящиеся к этому вопросу.

Рассмотрим сначала сегнетоэлектрик типа порядок—беспорядок с симметричным двухминимумным одночастичным потенциалом и одним диполем на ячейку (например, $KH_2PO_4$). В псевдоспиновом представлении [79] гамильтониан такого сегнетоэлектрика, взаимодействующего с одномодовым полем, имеет вид

$$\mathcal{H} = -\hbar\Omega \sum_{i=1}^{N} \sigma_i^x + \hbar\omega a^+ a - \frac{1}{2} \sum_{i,j} \mathcal{J}_{ij} \sigma_i^z \sigma_j^z + \frac{i\lambda}{\sqrt{N}} \sum_{i=1}^{N} \sigma_i^z (a^+ - a),$$

где $\sigma_i^\alpha$ — матрицы Паули; $a^+$, $a$ — операторы рождения и уничтожения фотонов (предполагается, что поле поляризовано вдоль сегнетооси); $\Omega$ — частота туннелирования; $\mathcal{J}_{ij}$ — потенциал взаимодействия диполей $i$- и $j$-й ячеек;

$\lambda = Zd(2\pi\hbar\omega N/V)^{1/2}$; $Z$ — заряд туннелирующего иона; $d$ — расстояние между ямами одночастичного потенциала; $V$ — объем резонаторной полости; $\omega$ — затравочная частота фотонов. Используя представление когерентных состояний для фотонов, в приближении молекулярного поля по взаимодействию $\mathcal{J}_{ij}$ [78] для свободной энергии $f$ на один диполь получим ($\beta = 1/\kappa T$)

$$\beta f = \beta\hbar\omega y^2 + \frac{1}{2}\beta\mathcal{J}_0 s^2 - \ln\left\{2\,\text{ch}\left[\frac{\beta}{2}\sqrt{(\hbar\Omega)^2 + (\mathcal{J}_0 s - 2\lambda y)^2}\right]\right\}, \quad \mathcal{J}_0 = \sum_j \mathcal{J}_{ij},$$

где $y^2 = \lim_{N\to\infty} \langle a^+ a \rangle / N$ — среднее число фотонов на диполь, $s \equiv \langle \sigma^z \rangle$ ($2Zds$ — дипольный момент ячейки), причем $y^2$ и $s$ определяются из уравнений

$$\frac{2\hbar\Omega}{\tilde{\mathcal{J}}_0}\sqrt{1 + \left(\frac{\tilde{\mathcal{J}}_0\omega}{\lambda\Omega}\right)^2 y^2} = \text{th}\left\{\frac{\beta\hbar\Omega}{2}\sqrt{1 + \left(\frac{\tilde{\mathcal{J}}_0\omega}{\lambda\Omega}\right)^2 y^2}\right\}, \quad s^2 = \left(\frac{\hbar\omega}{\lambda}\right)^2 y^2, \quad (31)$$

где $\tilde{\mathcal{J}}_0 = \mathcal{J}_0 + 2\lambda^2/\hbar\omega$. При выполнении условия

$$2\hbar\Omega/(\mathcal{J}_0 + 2\lambda^2/\hbar\omega) < 1 \quad (32)$$

уравнение (1) имеет решение $y^2 \neq 0$ при $T < T_c^{(1)} = \hbar\Omega[2k\,\text{arcth}(2\hbar\Omega/\tilde{\mathcal{J}}_0)]^{-1}$.

При $T < T_c^{(1)}$ $y^2 = s = 0$. Таким образом, при выполнении обычного критерия сегнетоэлектрического перехода типа порядок—беспорядок $2\hbar\Omega/\mathcal{J}_0 < 1$ условие конденсации фотонов (32) выполняется независимо от величины $\lambda^2/\hbar\omega$.

Таким образом, при температурах ниже критической температуры фазового перехода в сегнетоэлектрическое состояние в системе устанавливается недиагональный дальний порядок по оптически активным переходам, возникающим вследствие туннелирования одного из ионов сегнетоэлектрической ячейки в одночастичном двухминимумном потенциале внутри ячейки (так называемый потенциал Лифшица). Отметим также, что в этом случае ячейки сегнетоэлектрика, в принципе, могут рассматриваться в качестве двухуровневых кубитов, причём диполь-дипольное взаимодействие между ними способствует когерентизации их рабочего перехода.

Рассмотрим в заключение сегнетоэлектрик типа смещения, для которого одним из возможных механизмов возникновения сегнетоэлектрического состояния является электрон-фононное взаимодействие, приводящее к межзонным переходам [79-82]. В простейшей модели двух узких зон без дисперсии и одномодового затравочного предельного оптического колебания решетки гамильтониан сегнетоэлектрика, взаимодействующего с одной резонаторной модой, принимает форму

$$\mathcal{H} = \sum_{\mathbf{k}} \frac{\varepsilon}{2}\left[c_\sigma^+(\mathbf{k}) c_\sigma(\mathbf{k}) - c_\mu^+(\mathbf{k}) c_\mu(\mathbf{k})\right] + \hbar\Omega_0 b^+ b + \hbar\omega_\mathbf{q} a_\mathbf{q}^+ a_\mathbf{q} + \left\{\gamma a_\mathbf{q}^+ b + \sum_{\mathbf{k}}\left[\frac{\lambda_1}{\sqrt{N}} c_\mu^+(\mathbf{k}) c_\sigma(\mathbf{k}) b^+ + \frac{\lambda_2}{\sqrt{N}} c_\mu^+(\mathbf{k}) c_\sigma(\mathbf{k}) a_\mathbf{q}^+\right]\right\} + \{\text{э. с.}\},$$
(33)

где $c_\sigma^+(\mathbf{k})$, $c_\sigma$ — операторы фононной моды (волновой вектор q, частота $\omega_q$); $\varepsilon$ — средняя разность энергий зон σ и μ, $\lambda_1$ — константа вибронного взаимодействия

$$\lambda_2 = -(\pi e\hbar/m^* l)(2\pi\hbar N/\omega_\mathbf{q} V)^{1/2}, \quad \gamma = i(\pi e^2 \hbar^2 \omega_\mathbf{q} N/M\Omega_0 V)^{1/2}(\mathbf{e_q n}),$$
(34)

$m_*$ и $e$ эффективная масса и заряд электрона; $M$ — масса иона решетки; $N$ — число кристаллических ячеек; $l$ — период решетки; $\mathbf{e_q}$, $\mathbf{n}$ — векторы поляризации фотонной и фононной мод. В (33), (34) использовано, что для оптических и инфракрасных частот $k+q \simeq k$. Свободная энергия имеет вид

$$-\beta f = -\beta \frac{\lambda_1^2}{\varepsilon} \delta_1 y_1^2 - \beta \frac{\lambda_2^2}{\varepsilon} \delta_2 y_2^2 + 2\beta \frac{\lambda_1 \lambda_2}{\varepsilon} \chi y_1 y_2 \sin\varphi + 2\ln\left\{2\,\text{ch}\left[\frac{\beta\varepsilon}{4}\sqrt{1+\left(\frac{2\lambda_1}{\varepsilon}\right)^2 y_1^2 + \left(\frac{2\lambda_2}{\varepsilon}\right)^2 y_2^2 + 8\frac{\lambda_1\lambda_2}{\varepsilon^2} y_1 y_2 \cos\varphi}\right]\right\},$$
(35)

где $\delta_1 = \varepsilon\hbar\Omega_0/\lambda_1^2$, $\delta_2 = \varepsilon\hbar\omega_\mathbf{q}/\lambda_2^2$, $\chi = \varepsilon|\gamma|/\lambda_1\lambda_2$; $y_1^2 = \lim_{N\to\infty}\langle b^+ b\rangle/N$ дает средний квадрат смещения ячейки, $y_2^2 = \lim_{N\to\infty}\langle a_q^+ a_q\rangle/N$ — среднее число фотонов на одну ячейку, причем $y_1$, $y_2$, $\varphi$ находятся из уравнений, определяющих точку перевала функции (35) [21]. В типичных случаях $\Omega_0 \approx 10^{13}$ сек.$^{-1}$, $M \simeq 10^{-22}$ г, $l \simeq 5$ Å, $|m^*| \simeq 10^{-27}$ г, $\varepsilon \simeq (0{,}05\text{-}3)$ эв, $N/V \lesssim 10^{22}$ см$^{-3}$ (так как $V$ может превышать объем кристалла). Отсюда, согласно (34), $\chi$, $\chi^2/\delta_1\delta_2$, $\chi^2/\delta_1^2 \ll 1$. Тогда из уравнений для точки перевала можно получить

(36)
$$\frac{\delta_1\delta_2}{\delta_1+\delta_2}\sqrt{1+\left(\delta_i\frac{\delta_1\delta_2}{\delta_1\delta_2}\right)^2\left(\frac{2\lambda_i}{\varepsilon}\right)^2 y_i^2}=$$
$$=\text{th}\left\{\frac{3\varepsilon}{4}\sqrt{1+\left(\delta_i\frac{\delta_1+\delta_2}{\delta_1\delta_2}\right)^2\left(\frac{2\lambda_i}{\varepsilon}\right)^2 y_i^2}\right\}, \quad i=1, 2.$$

При условии

(37)
$$(\lambda_1^2/\varepsilon\hbar\Omega_0)+(\lambda_2^2/\varepsilon\hbar\omega_q)>1$$

уравнения (36) описывают конденсацию фотонов и фононов при $T<T_c^{(2)}=$
$=\varepsilon\{4k \text{ arc th }[(\delta_1^{-1}+\delta_2^{-1})^{-1}]\}^{-1}$. При $T>T_c^{(2)}$: $y_1^2=y_2^2=0$. Если выполнен критерий сегнетоэлектрического перехода $\lambda_1^2/\hbar\varepsilon\Omega_0>1$, то (7) удовлетворяется при любой величине $\lambda_2$.

Необходимо отметить, что значительное внимание было уделено нами исследованию спектра элементарных возбуждений, в частности, вопросу о связи фазового перехода с существованием возбуждений типа мягкой моды и исследованию влияния поля на характер равновесного состояния кристалла. Это исследование и его результаты будут приведены в последующих публикациях; пока же укажем, что исследование спектра элементарных возбуждений в равновесной системе «сегнетоэлектрик (антисегнетоэлектрик) + электромпгнитное поле» показывает, что фазовый переход, обусловливающий конденсацию фотонов, связан с существованием коллективных возбуждений типа мягкой моды. Характер «размягчения» этой моды (то есть поведение частоты при температурах, близких к критической) в главных чертах не отличается от того, который имеет место для мягкой моды обычного сегнетоэлектрика

Итак, для обоих типов сегнетоэлектриков критерии конденсации фотонов (32) и (37) выполняются независимо от величины константы взаимодействия с полем. Последняя влияет только на равновесное число заполнения резонаторной моды. Критические температуры конденсации $T_c^{(1)}$ и $T_c^{(2)}$ превышают температуры сегнетоэлектрических переходов. Следовательно, рассмотренные системы могут дать возможность получения эффекта конденсации фотонов в реальных условиях (различные его проявления рассмотрены в [11, 15, 16, 62, 63, 64]).

В заключение необходимо еще раз отметить, что материал раздела 4, касающиеся особенностей возникновения недиагонального дальнего порядка (когерентизации всей системы) в равной степени относится как к ансамблю атомов или молекул, так и к ансамблю реальных спинов, взаимодействующих с переменным электромагнитным полем. В частности, он справедлив для ферромагнетика с поперечным взаимодействием между спинами.

## Заключение

В недавних работах [1-6] нами был предложен новый способ формирования сети кубитов на основе самоорганизующихся систем. В связи с этим в настоящей работе проведено теоретическое исследование некоторых из самоорганизующихся систем, в которых могут возникать упорядоченные структуры, обладающие свойствами, перспективными для организации такой сети.

В начале статьи изложены, в частности, исходные положения и качественные соображения относительно потенциальных возможностей использования таких самоорганизующихся структур.

Далее проведено детальное теоретическое исследование тех самоорганизующихся структур, как диссипативных, так и равновесных, которые могут оказаться полезными для решения проблем масштабирования, устойчивости к декогерентизации и ошибкам, проблемы редуцирования негативного влияния температуры (в особенности на ансамблевые квантовые компьютеры) и других. При этом основное внимание было уделено изложению новых результатов, касающихся обоснования именно тех свойств, особенностей и характеристик самоорганизующихся систем, которые могут представлять интерес для физической реализации квантовых компьютеров.

Кратко перечислим основные результаты, относящиеся к диссипативных и равновесным самоорганизующихся структурам.

### *Диссипативные самоорганизующиеся структуры.*

В общем случае электромагнитного поля с произвольной длиной волны дана теория открытых резонаторов с периодическими границами (ОРПГ). На основе решения краевой задачи уравнений электродинамики проведено исследование типов собственных колебаний (мод) ОРПГ, вычислены поля внутри и вне резонатора для различных периодических структур на границах. Показано, что, как и в оптическом диапазоне, при определенных расстояниях между зеркалами резонатора, возникает специфическая структура пространственного распределения амплитуды поля (диссипативная самоорганизующаяся структура поля). Выяснены условия реализации таких периодических мод, подробно исследовано их пространственная структура, состоящая из периодического чередования резких δ-образных максимумов (то есть, в пределе больших отношений апертуры резонатора $A$ к периоду коэффициента пропускания $p$ интенсивность этих максимумов неограниченно возрастает с уменьшением их ширины). Обсуждены особенности и перспективы использования таких мод для формирования системы кубитов в квантовых компьютеров.

Из полученных результатов, в частности, следует оценка возможной степени масштабирования при использовании резонаторных систем. Так, число ловушек, когерентно связанных между собой периодической модой резонатора длины $L$ с апертурой $A$, будет порядка $8A^2L/p^3$, так как $(4A/p)^2$ – число узлов поля в плоскости, перпендикулярной оси резонатора, $2L/p$ – число таких плоскостей, расположенных вдоль оси резонатора и отстоящих друг от друга на расстоянии $p/2$ (см. выражения (13), (14) или (28). При этом длина резонатора должна быть равна

$$L = (2p^2/\lambda)n, \qquad (38)$$

где $n$ – произвольное натуральное число. Например, уже при числе периодов коэффициента пропускания на апертуре $A/p \sim 30$ и $L/p \sim 10$, число узлов (ловушек) будет порядка $10^4$. Из структуры стоячих волн (13), (14) или (28) следует также, что для рассмотренного резонатора с бесконечными зеркалами и при точном выполнении условий (38) ловушки будут иметь бесконечную глубину. При конечных размерах зеркал их глубина также может быть значительной, но конечной; оценка её величины требует отдельного исследования, в частности, учета эффекта туннелирования.

Отметим ещё, что, согласно результатам разделов 2, 3, меняя скважность периодической границы и длину резонатора, можно регулировать расположение узлов и пучностей самоорганизующейся диссипативной структуры поля, что важно при использовании ОРПГ для формирования сети кубитов; при этом, конечно, необходимо соблюдение условия (38) на длину резонатора.

Исследованы свойства резонаторных систем типа ОРПГ как диссипативной самоорганизующейся структуры, формирующейся при изменении длины резонатора; показана специфика эффекта кристаллизации поля и возникновения периодических мод с точки зрения теории фазовых переходов.

*Равновесные самоорганизующиеся структуры*.

В целях анализа равновесных структур, перспективных для формировании сети кубитов, проведено детальное исследование термодинамических свойств различных систем, состоящих из частиц с дискретным ограниченным спектром (энергетические или реальные спины) и поля излучения. Выяснены условия осуществления в таких системах фазового перехода второго рода, приводящего при температурах ниже критической к макроскопическому заполнению резонансной моды поля (к конденсации фотонов).

На примере системы «двухуровневые частицы — поле» выяснен физический смысл фазового перехода и условия его осуществления (условия сильной связи). С помощью представления когерентных состояний момента показана термодинамическая эквивалентность двухуровневой системы модели с прямым резонансным диполь-дипольным взаимодействием между частицами (аналогичной также XY-модели ферромагнетизма). В термодинамическом пределе исследован энергетический спектр системы. С помощью анализа энергетического спектра выяснено, что основным состоянием системы является глауберовское когерентное состояние но полю, амплитуда которого при температурах ниже критической пропорциональна $\sqrt{N}$ (*N* —число частиц), и когерентное состояние момента но суммарному энергетическому спину частиц.

Развит метод исследования термодинамических свойств различных многоуровневых систем, взаимодействующих с многомодовым полем, что существенно расширяет возможности исследования самоорганизующихся структур в формировании квантово-компьютерных систем. Так, показана возможность конденсации фотонов высокоэнергетического («замыкающего») перехода без выполнения условия сильной связи по этому переходу. Например, что если условие сильной связи выполняется по низкоэнергетическим переходам *l* ↔ *l—1* (*l* = 1,2, . . ., ***n;*** *1* — номер уровня), то для макроскопического заполнения моды, резонансной замыкающему переходу 1↔***n,*** выполнения этого условия уже не требуется, и возникает недиагональный дальний порядок по высокоэнергетическому.

Выявлены другие основные особенности поведения многоуровневых систем. В частности, получены условия совместного или раздельного заполнения различных мод поля, значения критических температур и критических индексов в зависимостях параметров порядка от температуры, выяснено поведение многоуровневых систем в области низких температур. Таким образом, данный тип самоорганизации может оказаться полезным для выполнения второго и третьего требований (обеспечение инициализации и блокирование декогерентизации, соответственно), которые сформулированы в работе DiVincenco *et al.* [83] как необходимые для реализации полномасштабного квантового компьютера.

Другой, важный для интересующих нас приложений, результат получен при исследовании фазовых превращений вещества, сопровождающихся возникновением недиагонального дальнего порядка (например, при сегнетоэлектрическом или ферромагнитном переходе). Показано, что, если прямое взаимодействие между частицами способно само приводить к фазовому переходу с недиагональным дальним порядком, то конденсация фотонов может происходить независимо от выполнения условия сильной связи по электромагнитному полю; другими словами, для получения макроскопического заполнения резонаторной моды электромагнитного поля выполнение этого условия не требуется. Таким образом, когерентизация системы частиц происходит вследствие того, что при температурах ниже критической в системе возникает недиагональный дальний порядок за счёт прямого взаимодействия частиц; тогда при наличии резонатора происходит конденсация фотонов, ответственных за радиационные переходы частиц между уровнями, на которые воздействует резонаторное поле.

Проведено детальное исследование термодинамического поведения двух моделей с прямым взаимодействием: сегнетоэлектрической (или ферромагнитной) системы типа порядок— беспорядок и типа смещения. В обоих случаях выявлены условия конденсации фотонов и показано, что фазовый переход в состояние с макроскопическим заполнением моды происходит одновременно с сегнетоэлектрическим (ферромагнитным — в случае реальных спинов) переходом, а выполнения условия сильной связи по полю в рассматриваемых системах не требуется. При этом конденсации фотонов повышает критическую температуру сегнетоэлектрического перехода.

Для системы типа порядок—беспорядок выяснено, что как в случае сегнетоэлектрического, так и в случае антисегнетоэлекрического упорядочения конденсация фотонов возможна, если волновой вектор поля равен волновому вектору зоны Бриллюэна, при котором достигает максимума фурье-компонента потенциала взаимодействия диполей кристаллических ячеек. Установлено, что фазовый переход, сопровождающийся конденсацией фотонов, связан с существованием в системе возбуждений типа мягкой моды. Исследованы и другие свойства такого рода самоорганизующихся структур, которые обладают определёнными своими преимуществами с точки зрения когерентизации кубитов.

Отметим, что для сегнетоэлектриков типа порядок – беспорядок при температурах ниже критической температуры перехода в сегнетоэлектрическое состояние в системе устанавливается недиагональный дальний порядок по оптически активным переходам, возникающим вследствие туннелирования одного из ионов сегнетоэлектрической ячейки в одночастичном двухминимумном потенциале внутри ячейки. В этом случае ячейки сегнетоэлектрика могут рассматриваться в качестве двухуровневых кубитов, причём оказывается, что именно их диполь-дипольное взаимодействие способствует когерентизации рабочего перехода. Представляется, что организация кубитовой сети на базе подобного рода систем с прямым взаимодействием, обеспечивающим ее когерентизацию, могла бы способствовать выполнению не только второго и третьего требований, сформулированных в работе [83] (см. выше), но также и четвертого требования (относящего к передача информации между кубитами).

В заключение отметим, что при исследовании самоорганизующихся структур, пригодных для целей формирования кубитовых систем, полезными могут оказаться и такие способы формирования сети кубитов, которые сочетают оба вида структур - как диссипативных, так и равновесных.

## ЛИТЕРАТУРА